\documentclass[twocolumn,pra,showpacs,superscriptaddress]{revtex4}
\usepackage{amssymb}
\usepackage{bm}
\usepackage{array}
\usepackage{graphicx}
\usepackage{caption2}
\usepackage{amsmath}
\usepackage{subfigure}
\usepackage[sort&compress]{natbib}

\begin{document}
  \title{Generating EPR beams in a cavity optomechanical system}
\author{Zhang-qi Yin}
  \affiliation{Department of Applied Physics, Xi'an Jiaotong University, Xi'an 710049, China}
  \affiliation{FOCUS center and MCTP, Department of Physics, University of Michigan, Ann Arbor, Michigan 48109, USA}
\author{Y.-J Han}
 \affiliation{FOCUS center and MCTP, Department of Physics, University of Michigan, Ann Arbor, Michigan 48109, USA}

 \begin{abstract}
We propose a scheme to produce continuous variable entanglement
between phase-quadrature amplitudes of two light modes in an
optomechanical system. For proper driving power and detuning, the
entanglement is insensitive with bath temperature and $Q$ of
mechanical oscillator. Under realistic experimental conditions, we
find that the entanglement could be very large even at room
temperature.
\end{abstract}
 \maketitle


Entanglement is the key resource of the field of quantum
information. Light is the perfect medium to distribute entanglement
among distant parties. Entangled light with continuous variable (CV)
entanglement between phase-quadrature amplitudes of two light modes
is widely used in teleportation, entanglement swap, dense coding,
etc. \cite{Braunstein05}. This type of entangled state is also
called Einstein-Podolsky-Rosen (EPR) state. The EPR beams have been
generated experimentally by a nondegenerate optical parameter
amplifier \cite{PhysRevLett.68.3663}, or Kerr nonlinearity in an
optical fiber \cite{PhysRevLett.86.4267}. The later one is simpler
and more reliable. The Kerr nonlinearity is used to generate two
independent squeezed beams. With interference at a beam splitter,
the EPR entanglement is obtained between output beams. However, Kerr
nonlinearity in fiber is very weak, which limits entanglement
between output beams.

It was found that strong Kerr nonlinearity appeared in an
optomechanical system consisting of a cavity with a movable boundary
\cite{hilico85,PhysRevA.49.4055,PhysRevA.49.1337}. Besides, the
single-mode squeezing could be made insensitive with thermal noise
\cite{PhysRevA.49.4055}, which makes the scheme very attractive.
However, the frequency of output squeezed beams cannot be made
identical, which makes interference difficult. Then it was
generalized to two-mode schemes in order to generate EPR beams
without interference
\cite{giovannetti01,mancini01,pirandola03,genes08,Wipf:arXiv0803.4001}.
However, they are either very sensitive to thermal noises
\cite{giovannetti01,mancini01,genes08,Wipf:arXiv0803.4001}, or
requiring ultrahigh mechanical oscillator Q $\sim 10^8$ to suppress
thermal noise effects \cite{pirandola03}, which is $2$ to $3$ orders
higher than the present available parameters \cite{regal108}. The
practical scheme to generate EPR beams in an optomechanical system
needs to overcome these problems.

In this paper we propose a practical scheme to produce EPR beams in
an optomechanical system, which consists of a whispering-gallery
mode(WGM) cavity with a movable boundary. We find that, similarly as
the single-mode scheme \cite{PhysRevA.49.4055}, the thermal noise in
the two-mode scheme can be greatly suppressed by adiabatically
eliminating an oscillator mode. By precisely tuning the laser power
and detuning, the oscillation mode is adiabatically eliminated and
two output sideband modes are entangled. Unlike the cavity-free
scheme \cite{pirandola03},  our scheme requires modest oscillator
$Q$. Besides, the output light is continuous in our scheme, other
than pulse in Ref. \cite{pirandola03}. The most attractive feature
of our scheme is that the entanglement between output beams is
nearly not changed under different bath temperature and $Q$ of the
mechanical oscillator. Within the experimentally available
parameters \cite{Kippenberg07,2008arXiv0805.1608S}, we find the
maximum two-mode squeezing could be higher than $16$ dB under room
temperature. The entanglement of formation (EOF) between two modes
is larger than $5$ \cite{PhysRevLett.91.107901}.
Since the coupling efficiency between cavity and fiber could be
larger than $99\%$ in the WGM cavity system
\cite{PhysRevLett.91.043902}, we neglect the coupling induced noises
in this paper.

\begin{figure}[htpb]
  \centering
  \includegraphics[bb=80 10 484 340,width=5cm]{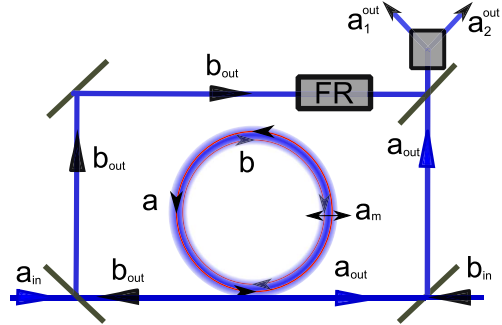}
  \caption{(Color online) Experimental setup. Cavity modes $a$ and $b$,
  which are driven by four lasers, couple to the mechanical mode $a_m$.}
  \label{fig:setup}
\end{figure}

As shown in Fig. \ref{fig:setup}, we consider an optomechanical
system consisting of a WGM cavity with a movable boundary. There are
two cavity modes $a$ and $b$ with the same frequency but the
opposite momentum. They are coupling with the same mechanical
oscillation mode $a_m$ and driven by four lasers, two from the
right-hand side with frequencies $\omega_L$ and $\omega_{L'}$, the
other two from the left-hand side with frequencies $\omega_L$ and
$\omega_{L'}$. $a_{\mathrm{in}}$ and $b_{\mathrm{in}}$ are input
lights. $a_1^{\mathrm{out}}$ and $a_2^{\mathrm{out}}$ are output
lights. Two lower mirrors have very high probability ($>99\%$) to
reflect the driving lasers. So we neglect the reflecting induced
noise for $a_\mathrm{out}$ and $b_\mathrm{out}$. The system
Hamiltonian is $H=H_0 +H_{d} + H_I$, where
\cite{2007PhRvL..99i3901W,2007PhRvL..99i3902M,2008Dayan,Srin07}
\begin{equation}
  \label{eq:Hamiltonian}
  \begin{aligned}
    H_0=&\hbar\omega_p (a^\dagger a +b^\dagger b)+  \hbar \omega_m
    a^\dagger_m a_m \\
    H_d=& \hbar (\frac{\Omega_a}{2}a + \frac{\Omega_b}{2} b) e^{-i\omega_L t} + \hbar
            (\frac{\Omega_a'}{2} a + \frac{\Omega_b'}{2} b)e^{-i\omega_L't} + \mathrm{H.c.}\\
    H_I= &\hbar \nu (a^\dagger b+ a b^\dagger)+ \hbar \eta \omega_m (a^\dagger a + b^\dagger b)
      (a_m + a^\dagger_m).
  \end{aligned}
\end{equation}
Here $a$, $b$ and $a_m$ are the annihilation operators for the
optical and mechanical modes, $\omega_p$ and $\omega_m$ are their
angular frequency. $\Omega_j$ with $j=a,b$ is the driving amplitude
and defined as $\Omega_j=2 \sqrt{P_j/\hbar \omega_L\tau}$, where
$P_j$ is the input laser power and $\tau=1/\gamma$ is the photon
loss rate into the output modes. $\nu$ is the coupling strength
between cavity modes $a$ and $b$. For the WGM cavity system, it
ranges from $100$ MHz to $10$ GHz \cite{2008Dayan,Srin07}. The
dimensionless parameter $\eta = (\omega_p/\omega_m) (x_m/R)$ is used
to characterize optomechanical coupling, with $x_m=
\sqrt{\hbar/m\omega_{m}}$ the zero-point motion of the mechanical
resonator mode \cite{2007OExpr..1517172K}, $m$ its effective mass,
and $R$ a cavity radius. In typical WGM cavity systems we find $\eta
\sim 10^{-4}$.

We define the normal modes $a_1= (a+b)/\sqrt{2}$ and
$a_2=(a-b)/\sqrt{2}$. We suppose the conditions that
$\Omega_a-\Omega_b=0$ and $\Omega_a'+ \Omega_b'=0$ are satisfied.
The Hamiltonian can be written as
\begin{equation}
  \label{eq:Hamiltonian1}
     \begin{aligned}
      H=&\hbar(\omega_p+\nu)a_1^\dagger a_1 + \hbar(\omega_p-\nu) a_2^\dagger a_2 +
 \hbar\omega_m a_m^\dagger a_m
        \\ &+\hbar(\frac{\Omega_1}{2} a_1 e^{-i\omega_Lt} + \frac{\Omega_2}{2} a_2 e^{-i\omega_L' t}
        + \mathrm{H.c.} )\\ &+ \hbar\eta \omega_m (a_1^\dagger a_1 + a_2^\dagger a_2) (a_m^\dagger + a_m),
    \end{aligned}
\end{equation}
where $\Omega_1= \Omega_a+ \Omega_b$ and $\Omega_2=
\Omega_a'-\Omega_b'$. We define the detuning $\Delta_1=
\omega_L-\omega_p-\nu$ and $\Delta_2= \omega_L'-\omega_p+\nu$. As
shown in Fig. \ref{fig:setup}, with beam splitters and Faraday
rotator, we can get the output mode of $a_1$ and $a_2$. We assume
both cavity and oscillator modes are weakly dissipating at rates
$\gamma$ and $\gamma_m$, respectively, where $\gamma_m\ll \omega_m$.
We can get quantum Langevin equations \cite{QO}
\begin{eqnarray}
  \label{eq:langevin1}
  \dot{a}_j &=& i\Delta_j a_j -i\eta \omega_m a_j (a_m+ a^\dagger_m) -
  i\frac{\Omega_j}{2} a_j-\frac{\gamma}{2}a_j(t)  \nonumber \\  &&+\sqrt{\gamma}
   a_j^{\mathrm{in}} ~~\mathrm{for} ~ j=1,2,\\
\label{eq:langevin2} \dot{a}_m &=& -i\eta\omega_m \sum_{j=1}^2 a_j^\dagger a_j -
(i\omega_m +\frac{\gamma_m}{2} ) a_m + \sqrt{\gamma_m} a_m^{\mathrm{in}},
\end{eqnarray}
where thermal noise inputs are defined as correlation functions
$\langle a_m^{\mathrm{in}\dagger}(t), a_m^{\mathrm{in}}(t') \rangle
= n_m \delta(t-t')$, $\langle a_m^{\mathrm{in}\dagger}(t),
a_m^{\mathrm{in}\dagger}(t') \rangle = \langle a_m^{\mathrm{in}}(t),
a_m^{\mathrm{in}}(t') \rangle = 0$, $\langle
a_j^{\mathrm{in}\dagger}(t), a_j^{\mathrm{in}}(t') \rangle =\langle
a_j^{\mathrm{in}\dagger}(t), a_j^{\mathrm{in}\dagger}(t') \rangle =
\langle a_j^{\mathrm{in}}(t), a_j^{\mathrm{in}}(t') \rangle = 0$,
with $n_m$ the thermal occupancy number of thermal bath for
oscillator mode. We suppose cavity modes couple with vacuum
bath. 

To simplify Eqs. \eqref{eq:langevin1} and \eqref{eq:langevin2}, we
apply a shift to normal coordinate, $a_j \rightarrow a_j+ \alpha_j$,
$a_m \rightarrow a_m + \beta$. $\alpha_j$ and $\beta$ are $c$
numbers, which are chosen to cancel all $c$ number terms in the
transformed equations. We find they should fulfill the following
requirements:
$    \beta
    \simeq -\eta(|\alpha_1|^2+ |\alpha_2|^2)$, and
$    i\Delta_j\alpha_j +2i\eta^2 \omega_m \alpha_j (|\alpha_1|^2+|\alpha_2|^2)
    - \frac{\gamma}{2} \alpha_j - i\frac{\Omega_j}{2}=0$.
Because $\gamma_m\ll \omega_m$, the imaginary part of $\beta$ can be neglected.
In the limit $\Delta_j \gg 2\eta^2w_m(|\alpha_1|^2+ |\alpha_2|^2)$, we find $\alpha_j
\simeq \Omega_j/\sqrt{\gamma^2+ 4\Delta_j^2}$.
 In the limit $|\alpha| \gg |\langle
a_p\rangle|$,  the Langevin equations are linearized as
\begin{eqnarray}
  \label{eq:langevin5}
  \dot{a}_j&=&  -i\eta\omega_m  \alpha_j(a_m+ a_m^\dagger) +(i\Delta_j'- \frac{\gamma}{2})a_j +
  \sqrt{\gamma} a_j^{\mathrm{in}}, \\
 \label{eq:langevin6}
    \dot{a}_m &=& -i\eta\omega_m \sum_{p=1}^2( \alpha_p^* a_p + \alpha_p a_p^\dagger
    ) - (i\omega_m + \frac{\gamma_m}{2}) a_m \nonumber \\ &&+ \sqrt{\gamma_m}
    a_m^{\mathrm{in}},
\end{eqnarray}
where $j=1,2$ and $\Delta_j'= \Delta_j +2 \eta^2 \omega_m
(|\alpha_1|^2+ |\alpha_2|^2)$. We suppose $\Delta_1'<0$ and
$\Delta_2'>0$. We define $\delta= (\Delta_2'-\Delta_1')/2-\omega_m$
and $d= -(\Delta_1'+\Delta_2')/2$.

In the limit $\omega_m \gg \delta, d, \gamma,\gamma_m$, the Langevin
equations \eqref{eq:langevin5} and \eqref{eq:langevin6} can be simplified as
\begin{equation}
  \label{eq:langevin7}
  \begin{aligned}
  \dot{a}_1=& -i d a_1 -i \eta \omega_m \alpha_1 a_m -\frac{\gamma}{2} a_1 +
  \sqrt{\gamma} a_1^{\mathrm{in}}, \\
  \dot{a}_2 =& -i d a_2 -i \eta \omega_m \alpha_2 a_m^\dagger - \frac{\gamma}{2} a_2 +
  \sqrt{\gamma}  a_2^{\mathrm{in}}, \\
  \dot{a}_m=& i \delta a_m -i\eta \omega_m (\alpha_1^* a_1 + \alpha_2 a_2^\dagger) -
      \frac{\gamma_m}{2}a_m +\sqrt{\gamma_m}a_m^{\mathrm{in}}.
  \end{aligned}
\end{equation}
With proper detuning and input power, we can always tune the cavity
mode amplitude $\alpha_1 =\alpha_2 = \alpha$. Define the Fourier
components of the intracavity field by $a(t) = \frac{1}{
\sqrt{2\pi}}
\int^\infty_{-\infty}e^{-i\omega(t-t_0)}a(\omega){d}\omega$. In the
limit $\delta \gg \omega, \gamma_m$,  we can adiabatically eliminate
the $a_m$ mode. We get $a_m (\omega)
\simeq \eta \frac{\omega_m}{\delta} (\alpha^* a_1 + \alpha
a_2^\dagger) - \frac{\sqrt{\gamma_m}}{i\delta} a_m^{\mathrm{in}}$.
Then we have quantum Langevin equations for $a_1(\omega)$ and
$a_2^\dagger(-\omega)$
\begin{equation}
  \label{eq:ap2}
\begin{aligned}
  -i \omega a_1(\omega) =& -ig' a_1(\omega) - ig a_2^\dagger(-\omega)
     -\frac{\gamma}{2}  a_1(\omega) \\&+ \sqrt{\gamma} a_1^{\mathrm{in}}(\omega)
  + \sqrt{\tilde{a}_m} a_m^{\mathrm{in}} (\omega), \\
  -i \omega a^\dagger_2(-\omega) =& ig' a_2^\dagger(-\omega) + ig a_1(\omega) -
  \frac{\gamma}{2} a_2^\dagger(-\omega) \\&+ \sqrt{\gamma} a_2^{\dagger\mathrm{in}}(-\omega)
  -\sqrt{\tilde{\gamma}_m} a_m^{\mathrm{in}}(\omega),
\end{aligned}
\end{equation}
where $g = \eta^2 \left|\alpha\right|^2 \omega_m^2/\delta$,
$\tilde{\gamma}_m = (\eta |\alpha| \omega_m/\delta)^2\gamma_m$, and
$g'= g+d$.  In Eq. \eqref{eq:ap2}, we neglect the phase of $\alpha$
because it is not important.

Denote $\vec{a}(\omega) =\binom{a_1(\omega)}{a_2^\dagger(-\omega)}$, $\vec{a}^{\mathrm{in}}
(\omega)= \binom{a_1^{\mathrm{in}}(\omega)}{a_2^{\mathrm{in}\dagger}(-\omega)}$ and
 $\vec{a}_m^{\mathrm{in}}(\omega)= \binom{a_m^{\mathrm{in}}(\omega)}{-a_m^{\mathrm{in}}(\omega)}$.
 We get the following matrix equation
\begin{equation}
  \label{eq:matrix}
  \mathbf{A} \vec{a}(\omega) = \sqrt{\gamma} \vec{a}^{\mathrm{in}}(\omega)
  + \sqrt{\tilde{\gamma}_m}\vec{a}_m^{\mathrm{in}},
\end{equation}
where
$$ \mathbf{A}=
\begin{pmatrix}
  -i\omega+ \frac{\gamma}{2}+ig'& ig\\
  -ig& -i\omega+ \frac{\gamma}{2}-ig'
\end{pmatrix}.
$$
Using boundary conditions $a_j^{\mathrm{out}} (\omega) = - a_j^{\mathrm{in}} (\omega) +
\sqrt{\gamma} a_j (\omega)$ for $j=1,2$, we can
calculate output field as
\begin{equation}
 \label{eq:output}
  \begin{aligned}
   a_1^{\mathrm{out}}(\omega)& =  G(\omega)
   a_1^{\mathrm{in}}(\omega)  - H(\omega)a_2^{\mathrm{in}\dagger}(-\omega) +
   I(\omega) a_m^{\mathrm{in}}(\omega) ,\\
   a_2^{\mathrm{out}\dagger}(\omega) &= G(\omega)^* a_2^{\mathrm{in}\dagger}(\omega)
   -H(\omega)^* a_1^{\mathrm{in}}(-\omega) - I(\omega)^* a_m^{\mathrm{in}}(-\omega),
  \end{aligned}
\end{equation}
where  $G(\omega)=
(\omega^2+ \frac{\gamma^2}{4}+ g^2 -g'^2- ig'\gamma)/\Delta(\omega) $, $H(\omega)=ig\gamma
/\Delta(\omega)$, $I(\omega)=(-i\omega+ \frac{\gamma}{2} -ig'+ig) \sqrt{\gamma
\tilde{\gamma}_m}/\Delta(\omega)$,
and $\Delta(\omega)= (-i\omega+\frac{\gamma}{2})^2 + g'^2-g^2$.

Let us define the dimensionless position and momentum operators of
fields $X_j^{\mathrm{out}} (\omega) = \big[ a_1^{\mathrm{out}}
(\omega) + a_1^{\mathrm{out}\dagger} (-\omega) \big]$ and
$P_j^{\mathrm{out}} (\omega) = \big[a_j^{\mathrm{out}}(\omega) -
a_j^{\mathrm{out}\dagger} (-\omega)\big]/i$, for $j=1,2$. We define
the correlation matrix of the output field as $V_{ij} = \langle
(\xi_i \xi_j + \xi_j\xi_i)/2\rangle$, where $\xi =
(X_1^{\mathrm{out}},P_1^{\mathrm{out}},X_2^{\mathrm{out}},P_2^{\mathrm{out}})$.
We calculate the correlation matrix with Eq. \eqref{eq:output}. Up
to local unitary transformation,
 the standard form of it is
\begin{eqnarray}
  \label{eq:correlation}
  V_S=
  \begin{pmatrix}
    n& 0& k_x& 0\\
    0& n& 0& -k_x\\
    k_x& 0& n& 0\\
    0& -k_x& 0& n&
  \end{pmatrix},
\end{eqnarray}
where $n=\{(\omega^2+\frac{\gamma^2}{4}+g^2-g'^2)^2 + (g'^2 +
g^2)\gamma^2 + [(\omega+g'-g)^2+\frac{\gamma^2}{4}]\gamma
\tilde{\gamma}_m(2n_m+1)\}/|\Delta (\omega)|^2$, $k_x =
\sqrt{V_{14}^2+V_{24}^2}$, where $V_{14} = -2g\gamma (w^2+
\frac{\gamma}{4} +g^2-g'^2) /|\Delta(\omega)|^2$, $V_{24} =
\{2g'g\gamma^2+[(\omega +g'-g)^2+\frac{\gamma^2}{4} ]\gamma
\tilde{\gamma}_m(2n_m+1)\}/|\Delta(\omega)|^2$. This is the
symmetric Gaussian state. The EOF for the symmetric Gaussian states
is defined as \cite{PhysRevLett.91.107901}
\begin{equation}
  \label{eq:EoF}
  \begin{aligned}
  E_F = C_+(n-k_x) \log_2[C_+(n-k_x)] \\ - C_-(n-k_x) \log_2[C_-(n-k_x)]
  \end{aligned}
\end{equation}
where $C_\pm (x)= (x^{-1/2} \pm x^{1/2})^2/4$. $V$ describes an
entanglement state if and only if $n-k_x<1$. Based on the standard
form of matrix \eqref{eq:correlation}, we also find that
$\langle\delta^2(X_1+X_2)\rangle= \langle \delta^2(P_1-P_2)\rangle =
n-k_x$. We define the two-mode squeezing as $S=-10 \log_{10}
(n-k_x)$.

 \begin{figure}[htbp]
   \centering
   \includegraphics[bb=109 263 484 572,width=7cm]{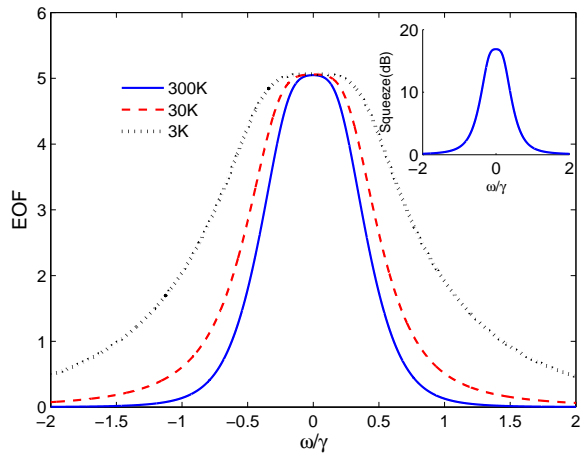}
   \caption{(Color online) EOF for different temperature.
   $|\alpha|$ is 1000, $\delta/2\pi=10$ MHz, $d=0.07\gamma$. The maxima squeezing
   is larger than $16$ dB when $T=300$ K.}
   \label{fig:temperature}
 \end{figure}
We now estimate the bath noise influence in experimentally
accessible conditions \cite{Kippenberg07}. The cavity resonant
frequency $\omega_p= 2\pi \times 300$ THz.
The oscillator frequency $ \omega_m =2\pi \times 73.5$ MHz. The
mechanical Q factor is about $30~000$. The cavity and oscillator
modes decay rates are $\gamma =2\pi\times 3.2$ MHz and $\gamma_m
=\omega_m/Q$, respectively. The cavity radius is $R=38$ $\mu$m. The
dimensionless coupling parameter is $\eta \simeq 10^{-4}$. We find
if $0<d\ll \gamma$, the thermal noise does not decrease the maximum
entanglement with strong enough input power. This is because we
adiabatically eliminate the mechanical mode and suppress the effects
of thermal noise. As show in Fig. \ref{fig:temperature}, the
temperature change does not change the maximum entanglement with
proper driving and detuning. But the higher the temperature, the
less the entanglement spectrum width. The embeded figure shows that
the maximum squeezing could be larger than $16$ dB at room
temperature.

 \begin{figure}[htbp]
   \centering
   \includegraphics[bb=109 263 484 567,width=7cm]{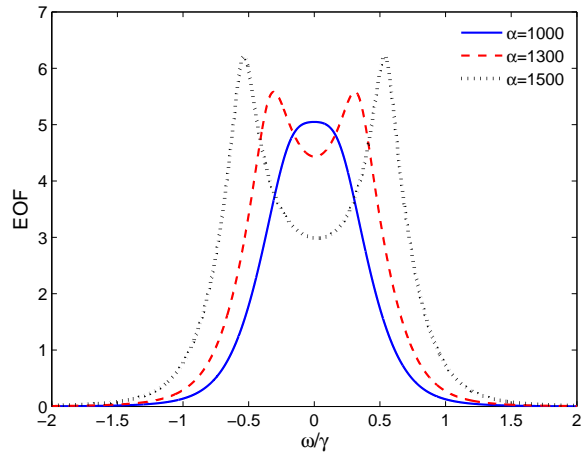}
 \caption{(Color online) EOF for different cavity mode amplitude $\alpha$.
 Here we adopt $\delta/2\pi=10$ MHz, $T=300$ K, 
 $\gamma_p/2\pi=3.2$MHz, and $d=0.07\gamma$. }
\label{fig:driving}.
 \end{figure}
As shown in Fig. \ref{fig:driving}, the bigger the cavity mode
amplitude $\alpha$, the larger the output entanglement. Because
$\alpha_j \simeq \Omega_j/\sqrt{\gamma^2+ 4\Delta_j^2}$, the output
entanglement is proportional to driving amplitude. But the peak of
entanglement is splitted into two symmetric peaks when driving is
very strong. The splitting distance is proportional to driving
power.
Increasing driving power can decrease the entanglement too. This is
because adiabatical elimination condition $\omega \ll \delta$ are
not valid around peaks for very strong driving.
 So the driving power should be neither too big nor too small.
For the specific $\alpha$ and $\delta$, we find there is an optimum
$d$ which makes entanglement maximum and the entanglement peaks
appear near $\omega=0$. The optimum $d$ is
$d_o=\sqrt{(\eta^2\omega^2 \alpha^2/\delta)^2+\gamma^2/4}
-(\eta\omega\alpha)^2/\delta$, corresponding to squeezing $S_o
=-10\log_{10}(4d_o^2/\gamma^2)$ and entanglement which is obtained
from Eq. \eqref{eq:EoF} with $n-kx=4(d_o/\gamma)^2$. It is obvious
that the higher the input power, the smaller the optimum $d$. In the
mean time, we find that decreasing the mechanical $Q$ factor nearly
does not change the entanglement spectrum if $d$ is around its
optimum value and the condition $ \omega_m/Q \ll \delta$ is
fulfilled. Leaving other parameters unchanged, $Q$ could be as low
as $300$. Considering the difficulty of increasing the mechanical
oscillator $Q$, the above finding makes our scheme more practical.

\begin{figure}[htbp]
   \centering
   \includegraphics[bb=109 263 484 567,width=7cm]{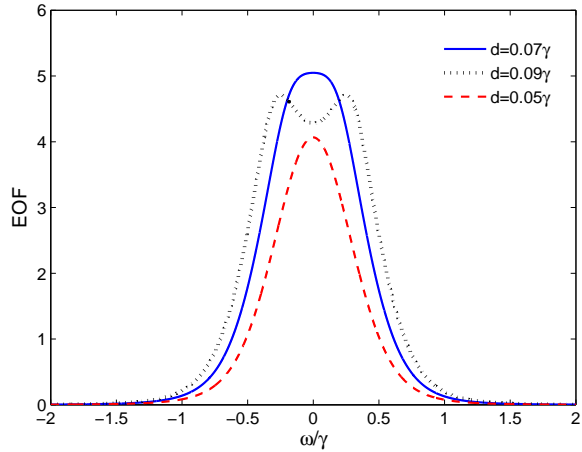}
 \caption{(Color online) EOF for different $d$.
 Here we adopt $\omega_m/2\pi=73.5$ MHz, $T=300$K, $\gamma_m=\omega_m/30~000$,
 $\gamma/2\pi =3.2$ MHz, and $|\alpha|=1000$. }
 \label{fig:driving2}.
 \end{figure}

We also test the stability of our scheme. As shown in Fig.
\ref{fig:driving2}, the optimum $d$ is around $0.07\gamma$ if
$\alpha=1000$, $\delta/2\pi=10$ MHz. To maintain such high
entanglement, we need to precisely control the $d$ down to
$0.02\gamma \sim 2\pi \times 60$ kHz. $d$ is defined as
$d=-(\Delta_1'+ \Delta_2')/2 = -(\Delta_1+\Delta_2)/2 -
4\eta^2\omega_m|\alpha|^2$. The higher entanglement is needed, the
more precise detuning and driving power is required at the same
time. To maintain the entanglement as high as Fig.
\ref{fig:driving2}, the laser spectrum width should be less than
$60$ kHz and the driving power fluctuation should be less than
$1\%$. The lower entanglement between two beams is needed to
maintain, the larger the optimum $d$ is. Therefore higher
fluctuations of detuning and driving power are allowed.

Before conclusion, we briefly discuss the approximations we used.
Our scheme needs the steady states existing, which requires $\langle
a_j^\dagger a_j\rangle \ll |\alpha|^2$. During numerical
calculation, $\langle a^\dagger a\rangle $ is in the order of
$10^3$, which is much less than $|\alpha|^2 \sim 10^6$. The other
two approximations are rotating wave approximation $\omega_m \gg
\delta, d, \gamma,\gamma_m$ and adiabatical elimination $\delta \gg
\omega, \gamma_m$, which can be fulfilled independently. For $\alpha
\sim 10^3$, the driving amplitude $\Omega$ is in the order of
$10^{11}$ Hz, which is much lower than the distance between adjacent
cavity modes $\Delta \omega =c/(Rn_0)\sim 5\times 10^{12}$ Hz, where
$c$ is the light speed in a vacuum, $n_0$ the refractive index of
silica. Therefore the approximation that one laser only drives one
cavity mode is valid. Laser power is needed in the order of $10$ mW,
which is available in the laboratory.


In conclusion, we have proposed a scheme to generate EPR lights in
an optomechanical system. Two sideband modes, which couple with the
mechanical mode, are driven by lasers. After adiabatically
eliminating the the mechanical mode, we find that the output
sideband modes are highly entangled. The higher power of the driving
laser, the larger entanglement of the output light. To maintain the
entanglement, we need to precisely control the driving power and
laser frequency at the same time. With proper parameters, the
entanglement is insensitive to the thermal noise and mechanical $Q$
factor. We test the scheme by experimental available parameters.
Though in this paper we fucus on WGM cavity systems, our scheme can
be realized in other optomechanical systems, as long as the
mechanical mode frequency is much larger than the cavity decay rate.

We thank Lu-ming Duan for helpful discussions. We thank Yun-feng
Xiao and Qing Ai for valuable comments on the paper. ZY was
supported by the Government of China through CSC (Contact
No.2007102530).


\end{document}